# Two Distinct Oxidation Dispersion Mechanisms in Pd-CeO$_2$ Mediated by Thermodynamic and Kinetic Behaviors of Single Pd Species

*Chen Zou,[1,#] Wen Liu,[1,#] Shiyuan Chen,[1,#] Songda Li,[1] Fangwen Yang,[1] Linjiang Yu,[1] Chaobin Zeng,[2] Yue-Yu Zhang,[3] Xiaojuan Hu,[1] Zhong-Kang Han,[1,\*] Ying Jiang,[1] Wentao Yuan,[1,\*] Hangsheng Yang,[1] Yong Wang[1,\*]*

[1]*Center of Electron Microscopy and State Key Laboratory of Silicon Materials, School of Materials Science and Engineering, Zhejiang University, Hangzhou, 310027, China.*
[2]*Hitachi High-Tech Scientific Solutions (Beijing) Co.,Ltd. Beijing, 100015, China.*
[3]*Wenzhou Institute, University of Chinese Academy of Sciences, Wenzhou, 325001, China.*
[#]*These authors contributed equally.*

Email: hanzk@zju.edu.cn; wentao_yuan@zju.edu.cn; yongwang@zju.edu.cn

Understanding the dispersion process of supported catalysts is crucial for synthesizing atomic-level dispersed catalysts and precisely manipulating their chemical state. However, the underlying dispersion mechanism remains elusive due to the lack of atomic-level evidence during the dispersion process. Herein, by employing spherical aberration-corrected environmental scanning transmission electron microscopy (ESTEM), first-principles calculations, and a global optimization algorithm, we unraveled the pre-oxidation dispersion and direct dispersion mechanisms in the Pd/CeO$_2$ (100) system, mediated by the thermodynamic and kinetic behaviors of single Pd species. We discovered that at lower temperatures, the Pd nanoparticles first undergo oxidation followed by the dispersion of PdO, while at higher temperatures, the entire dispersion process of Pd remains in a metallic state. The distinct dispersion mechanisms at different temperatures are driven by the thermodynamic and kinetic differences of environment-dependent single Pd species. The nonmobile Pd$_1$O$_4$ species stabilized at lower temperatures obstructs the direct dispersion of Pd nanoparticles, instead triggering a sequence of pre-oxidation followed by limited dispersion. In contrast, the highly mobile Pd$_1$O$_2$ species at higher temperatures facilitates the complete and direct dispersion of Pd nanoparticles. This research illuminates the essential physical mechanisms of oxidative dispersion from both thermodynamic and kinetic perspectives, potentially enabling strategies for precisely controlling the state of highly dispersed catalysts.

In heterogeneous catalysts, the size and the chemical state of metal nanoparticles significantly impact the activity and selectivity of supported catalysts [1–3]. Atomic-level dispersed catalysts are extensively studied in heterogeneous catalysis research due to their excellent catalytic performances in oxidation, hydrogenation, electrocatalysis, and photocatalysis [4–7]. Various methods, such as wet-chemistry synthesis, spatial confinement strategy, defect design strategy, chemical etching, and atomic layer deposition, have been developed to precisely fabricate the atomic-level dispersed catalysts through both chemical and physical routes [8,9]. Among these methods, the annealing process in specific atmospheres is widely used to develop atomic-level dispersed catalysts from supported transition metal nanoparticle catalysts like Pd, Pt, Rh, and Ru [10–13]. However, achieving atomically precise control over the size and chemical state of these dispersed catalysts remains challenging due to the complex factors involved in the annealing process. For instance, during the oxidation annealing process [14,15], not only does the competition between sintering and ripening occur [16–21], but other physical and chemical processes such as oxidation and phase transition [18,22,23] can also take place in high-temperature oxidation environments. Therefore, an in-situ study that decouples the influences of these complex factors and comprehensively understands the dispersion at the atomic level is highly desired. Such a study could potentially facilitate the design of high-performance catalysts and the regeneration of catalysts [24,25].

The Pd/CeO$_2$ catalyst is widely used in automobile exhaust treatment, methane oxidation, CO$_2$ reduction, and organic synthesis [26–29]. Ceria-supported atomic-level dispersed Pd has demonstrated exceptional activity and selectivity across various reactions [11,30]. Numerous efforts have been made to understand and precisely control the dispersion of Pd/CeO$_2$ catalysts at the atomic level, including their coordination environment and chemical state [4,31–33], although the state of the highly dispersed Pd atom/cluster is very sensitive to environmental conditions [34,35]. Through theoretical approaches, Hu and Li revealed that the Sabatier principle could elucidate the dispersion and sintering phenomena of supported metal nanoparticles via theoretical calculations [36]. In practical applications, however, the dispersion and sintering processes are notably intricate, involving the oxidation and reduction of supported Pd species [37]. For instance, Jiang *et al*. recently discovered that the coordination numbers of Pd atoms in Pd/CeO$_2$ catalysts could be affected by the gas environment, which, in turn, influences the dispersion and sintering behavior of Pd species [31]. Additionally, Muravev *et al*. demonstrated that altering the size of the ceria support adjusts oxygen mobility and metal-support interactions, thereby influencing the dispersion and activity of Pd species [33]. However, to unravel the dispersion mechanisms and achieve controllable manipulation of highly dispersed noble metal catalysts, detailed atomic-scale information during the dispersion process is highly desired.

In this work, we in-situ investigated the atomic-level dynamic dispersion behavior of Pd/$CeO_2$ catalysts across various temperatures under an $O_2$ atmosphere using environmental scanning transmission electron microscopy (ESTEM), complemented by first-principles calculations and a global optimization algorithm. The in-situ experiments revealed two distinct dispersion processes of Pd species on the $CeO_2$ (100) surface: (1) pre-oxidation dispersion and (2) direct dispersion, which occurred at different temperatures in the oxygen environment. The calculations uncovered temperature-dependent dispersion mechanisms governed by the stabilities and mobilities of single Pd species on the $CeO_2$ (100) surface. This work provides a comprehensive view of the dispersion mechanisms from both thermodynamic and kinetic perspectives, which can help to precisely adjust the catalyst size and coordination number, thereby promoting the development of efficient catalysts.

To investigate the dispersion behavior of Pd/$CeO_2$ catalysts, $CeO_2$ nanocubes were synthesized with uniform morphology using the hydrothermal method [38], and Pd nanoparticles were subsequently loaded through the solid grinding method [39]. For detailed synthesis methods, please refer to the Supporting Information. As shown in Fig. 1a, Pd nanoparticles are uniformly distributed on the surface of the $CeO_2$ support, with an average size of approximately 4.34 ± 0.86 nm (Fig. 1g). The high-resolution image in Fig. 1c shows that the interplanar spacing of the Pd nanoparticles is 0.22 nm, corresponding to the (111) surface. Fig. 1b displays the condition of fresh Pd/$CeO_2$ after being treated at 823 K in air for 4 hours, where a notable reduction in Pd particle size is observed (Fig. 1h), along with a significant increase in the number of clusters on the $CeO_2$ surface. To further confirm the dispersion of Pd on the $CeO_2$ support, secondary electron (SE) images were captured during in-situ ESTEM observations, as shown in Fig. 1e and Fig. 1f. After 80 minutes of treatment at 773 K under 5 Pa $O_2$ conditions, a noticeable reduction in the size of Pd particles on $CeO_2$ was observed without the formation of larger Pd particles, indicating effective dispersion rather than sintering and growth. These observations align with previous findings that Pd/$CeO_2$ disperses into Pd atom catalysts in an oxygen atmosphere and undergoes sintering in a reducing atmosphere.

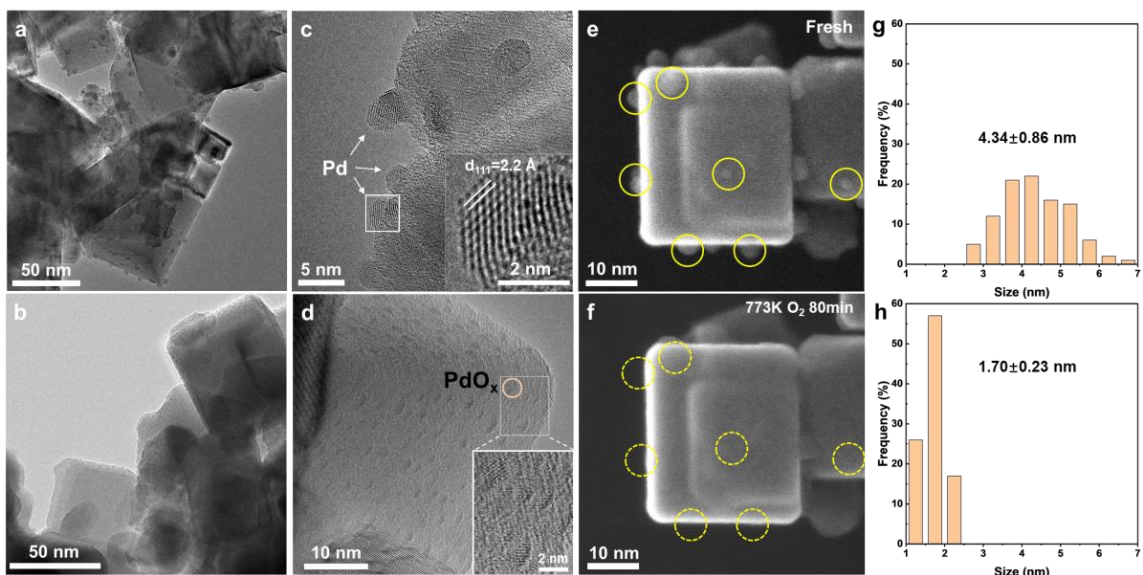

FIG. 1. Structures and morphologies of the Pd/CeO$_2$-nanocube catalysts. (a), (b) TEM images of Pd/CeO$_2$-cube and Pd/CeO$_2$-cube-O$_2$. (c), (d) HRTEM images of Pd/CeO$_2$-cube and Pd/CeO$_2$-cube-O$_2$; insets: a larger view of the particles. (e), (f) Low magnification SE images of in situ oxidative dispersion process. (g), (h) Particle size distributions of Pd NPs in Fig. 1a and Fig. 1b.

The in-situ dispersion experiment was first performed in a vacuum (10$^{-5}$ Pa) at 473 K to prevent organic contamination. Under these conditions, it was confirmed that Pd retained its metallic state, as shown in Fig. 2a. The Pd nanoparticle, exhibiting the (111) and (200) surface, formed an epitaxial interface with the (100) surface of CeO$_2$. Upon elevating the temperature to 673 K and introducing 5 Pa of O$_2$, the metallic Pd was fully oxidized to PdO, as shown in Fig. 2b. This oxidation expanded the contact interface with CeO$_2$, a change attributed to the favorable lattice match between the (002) surface of PdO and the (100) surface of CeO$_2$.

The dispersion of PdO nanoparticles began at the PdO-CeO$_2$ interface and progressed layer by layer along the (101) surface, as evidenced in Fig. 2c. Profile intensity analysis was conducted on the two atomic layers delineated by orange and green boxes (Fig. 2k-n). Initial dispersion was observed at the contact point of the first layer on the left side, indicated by the arrow in Fig. 2k, and subsequently occurred on the right side's first layer (Fig. 2d and 2m). Despite the slow rate of dispersion and its limited extent, it is suggested that the PdO nanoparticles may undergo self-adjustment to maintain thermodynamic stability during this process. As the holding time progressed, dispersion began consistently from the apex of the first layer on the left side (Fig. 2e) and eventually led to the complete dispersion of the first layers on both sides (Fig. 2f). Fig. 2k and 2m illustrate that over time, only a gentle curve remained, indicating the complete dispersion of the first layer of atoms with subsequent layers beginning to disperse. Overall, this layer-by-layer dispersion significantly reduced the vertical size of the PdO nanoparticles and decreased their contact area with $CeO_2$.

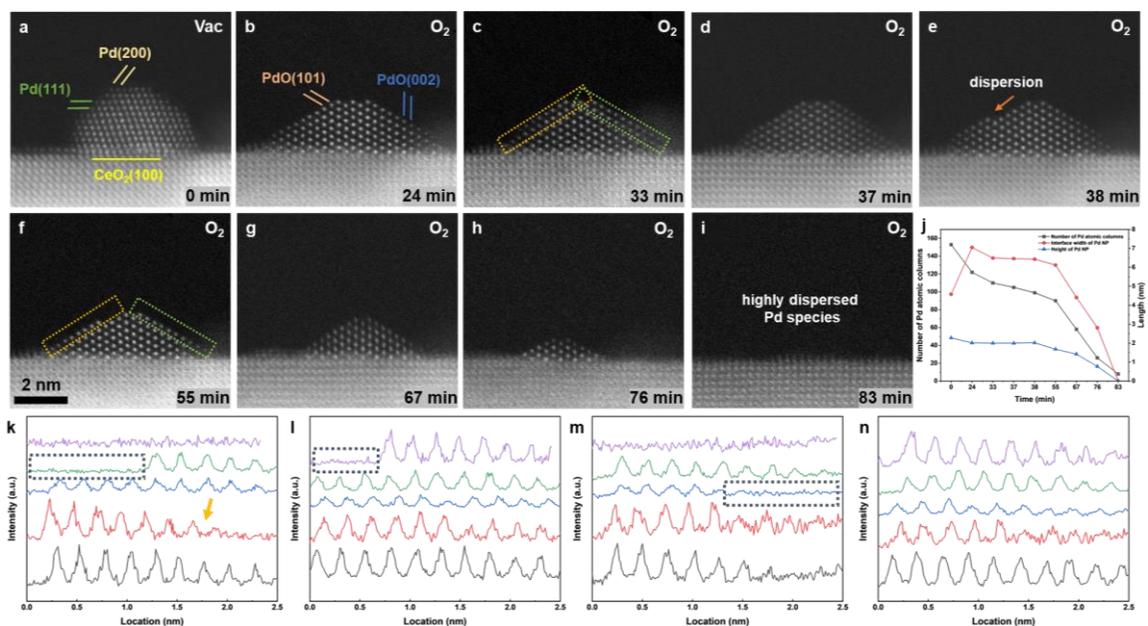

FIG.2. In situ oxidative and dispersion process of Pd NPs supported on the $CeO_2$-cube surface. (a), (b) ADF STEM images showing the oxidation of Pd on $CeO_2$-cube (100) surface in 5 Pa oxygen. (c)-(i) Time-sequenced ADF STEM images showing the dispersion of PdO NPs. The corresponding temperature of each picture is 473 K (a), 673 K (b)-(c), 723 K (d)-(f), and 773 K (g)-(i), respectively. (j) Changes in the number of atomic columns during the Pd dispersion process, the contact width between Pd or PdO nanoparticle and $CeO_2$, and the height of the Pd nanoparticle. (k)-(n) In Fig. 2b-2f, the profile intensity changes of the first (k) and second (l) layer atoms in the orange box and the first (m) and second (n) layer atoms in the green box. (The lines from bottom to top represent Fig. 2b-f respectively).

To accelerate the dispersion process, the temperature was raised to 773 K. A comparison between Fig. 2f and 2g revealed a further reduction in the size of the PdO nanoparticles, although they retained their triangular morphology. This size reduction occurred simultaneously with rapid dispersion, maintaining the layer-by-layer dispersion characteristic of the PdO particles, ultimately leaving a residual layer of Pd atoms on the surface. Fast Fourier Transform (FFT) patterns (Fig. S1), confirm the sequence of Pd oxidation to PdO prior to dispersion. This behavior was consistent across varying particle sizes, as evidenced in smaller (Fig. S2) and larger particles (Fig. S4), with the latter requiring a longer dispersion time.

The in-situ dispersion experiment was also performed at a high temperature. The initial Pd particle, primarily exposing the (200) and (111) surfaces (Fig. 3a), was directly heated to 773 K with the introduction of 5 Pa of $O_2$. A noticeable reduction in the size of the Pd particle was observed within just 4 minutes (Fig. 3b). Unlike the results at lower temperatures, despite this size reduction, the particles retained the Pd lattice structure instead of forming PdO. As time progressed, the lateral dimensions of the Pd particle decreased first, followed by a reduction in height. Yet, no lattice expansion occurred, indicating no PdO formation throughout the process. Notably, compared to the low-temperature process involving initial oxidation followed by dispersion, which took 83 minutes, the high-temperature dispersion without oxidation was completed in just 8 minutes. The corresponding FFT patterns (Fig. S7) further confirm that Pd dispersed without forming PdO. This behavior was consistent across other particles as well (Fig. S8). This indicates that the process of oxidative dispersion does not necessarily involve the formation of oxides, challenging traditional assumptions about the behavior of Pd particles under oxidative conditions.

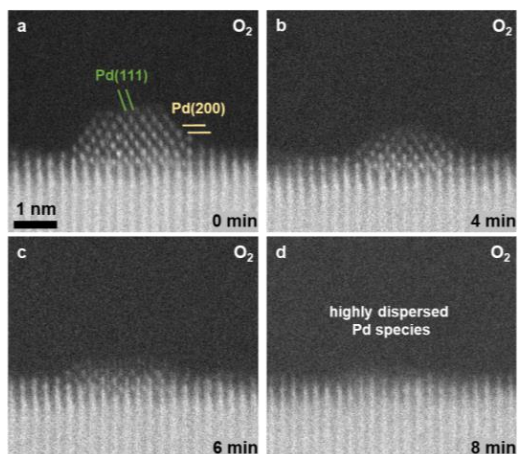

FIG. 3. (a)-(d) In situ observation of dynamic dispersion of a Pd NP supported on the $CeO_2$-cube surface under 773 K and 5 Pa $O_2$. The corresponding temperature of each picture is the same as 773 K.

X-ray Photoelectron Spectroscopy (XPS) was used to examine the valence states of Pd species. As shown in Fig. S12, the initial Pd/CeO$_2$ sample predominantly exists in the metallic state (Pd$^0$). In contrast, after treatment at 673 K, the sample predominantly shows Pd in the +2 oxidation state (Pd$^{2+}$). Following the 773 K treatment, it mainly displays Pd species with valence states ranging between Pd$^{2+}$ and Pd$^{4+}$. The observed changes in valence states corresponding to the treatment temperatures can be linked to the dispersion level of Pd. A higher valence state of Pd indicates a more extensive degree of dispersion, which is consistent with the TEM results (Fig. S13).

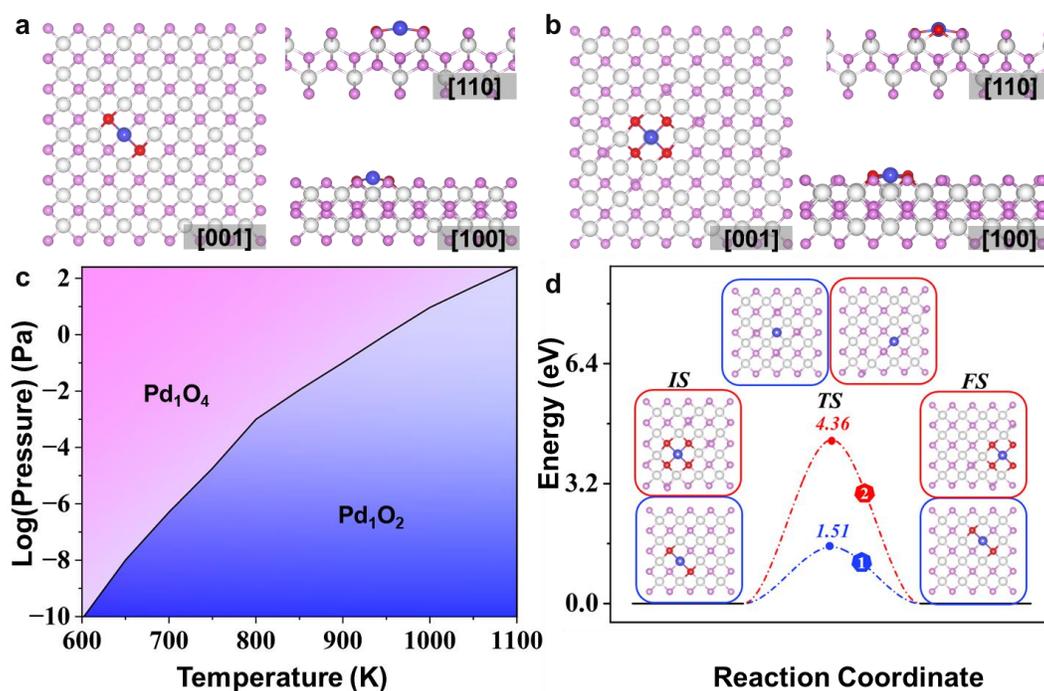

FIG. 4. Configurations of Pd$_1$O$_2$ (a) and Pd$_1$O$_4$ (b) motifs on CeO$_2$(100) surface. Pd, O, and Ce atoms are shown in blue, purple, and white, respectively. The O atoms connected to Pd are highlighted in red. (c) Phase diagram of Pd$_1$O$_2$ and Pd$_1$O$_4$ on the CeO$_2$(100) surface. (d) Energy profile for the diffusion pathways of Pd$_1$O$_2$ (blue line) and Pd$_1$O$_4$ (red line) on the CeO$_2$(100) surface.

To further elucidate the mechanisms driving the distinct dispersion behaviors observed at different temperatures, we conducted first-principles calculations and global structure searching (for detailed methodology, please refer to the Supplementary Methods section). This analysis integrates both thermodynamic and kinetic perspectives to enhance our understanding of the underlying processes. In our global structure searching, we explored a series of Pd-O-Ce configurations with various numbers of Pd and O atoms (Fig. S15). Our findings reveal that under oxidative conditions, single-atom Pd species are thermodynamically more stable than Pd clusters. Moreover, the single-atom Pd exhibits two motifs, $Pd_1O_2$ and $Pd_1O_4$, as shown in Fig. 4a and 4b. The $Pd_1O_4$ motif is found to be more stable than the $Pd_1O_2$ at lower temperatures in oxidative conditions (Fig. 4c), and it presents a high diffusion energy barrier of 4.36 eV along the (100) direction. Other potential diffusion pathways are also considered (Fig. S16). This high diffusion energy barrier suggests that the detachment of single-atom Pd from Pd nanoparticles into the $Pd_1O_4$ motif is inhibited when the Pd nanoparticles are surrounded by $Pd_1O_4$. To facilitate further dispersion of Pd nanoparticles, it is necessary to increase the temperature to transform $Pd_1O_4$ into $Pd_1O_2$, which significantly reduces the diffusion energy barriers to approximately 1.5 eV along the (100) directions. These computational insights are consistent with experimental observations at lower temperatures, where the dispersion rate is slow, and the extent of dispersion is limited. Instead of dispersing, Pd nanoparticles tend to oxidize into PdO nanoparticles at lower temperatures due to the higher thermodynamic stability of PdO (Fig. S17) compared to metallic Pd, a finding that aligns with experimental evidence. Conversely, at higher temperatures, the $Pd_1O_2$ motif becomes more stable compared to $Pd_1O_4$, and oxidized Pd nanoparticles become thermodynamically less stable as temperature increases, promoting a tendency for Pd metal nanoparticles to disperse directly into single atoms within the $Pd_1O_2$ motifs. The lower diffusion energy barrier for $Pd_1O_2$ motifs facilitates this process, leading to experimental observations at higher temperatures where direct dispersion occurs without the formation of PdO nanoparticles. Therefore, the state of the Pd single atoms can be controlled by dispersion at different temperatures.

In summary, we utilized in-situ spherical aberration-corrected environmental scanning transmission electron microscopy, first-principles calculations, and global optimization algorithms to elucidate the pre-oxidation dispersion and direct dispersion mechanisms of $Pd/CeO_2$ catalysts at varying temperatures. At lower temperatures, single-atom Pd predominantly exists as $Pd_1O_4$, characterized by a high diffusion energy barrier that restricts the dispersion of Pd nanoparticles and results in the formation of PdO nanoparticles. In contrast, at higher temperatures, the $Pd_1O_2$ motif becomes more stable relative to $Pd_1O_4$, and the oxidized Pd nanoparticles become thermodynamically less stable as temperature increases. This shift facilitates the direct dispersion of Pd metal nanoparticles into single atoms within the $Pd_1O_2$ motifs. This comprehensive understanding of the thermodynamic and kinetic properties of various supported Pd species under different conditions highlights the complex nature of Pd nanoparticles in catalytic systems and provides vital insights into optimizing catalytic activity and stability.

**REFERENCE**

[1] M. Valden, X. Lai, and D. W. Goodman, Science **281**, 1647 (1998).
[2] E. Roduner, Chem. Soc. Rev. **35**, 583 (2006).
[3] H. Wang, X.-K. Gu, X. Zheng, H. Pan, J. Zhu, S. Chen, L. Cao, W.-X. Li, and J. Lu, Sci. Adv. **5**, eaat6413 (2019).
[4] A. Wang, J. Li, and T. Zhang, Nat. Rev. Chem. **2**, 65 (2018).
[5] S. K. Kaiser, Z. Chen, D. Faust Akl, S. Mitchell, and J. Pérez-Ramírez, Chem. Rev. **120**, 11703 (2020).
[6] L. Liu and A. Corma, Chem. Rev. **118**, 4981 (2018).
[7] R. Li, L. Luo, X. Ma, W. Wu, M. Wang, and J. Zeng, J. Mater. Chem. A **10**, 5717 (2022).
[8] S. Ji, Y. Chen, X. Wang, Z. Zhang, D. Wang, and Y. Li, Chem. Rev. **120**, 11900 (2020).
[9] Y. Chen, S. Ji, C. Chen, Q. Peng, D. Wang, and Y. Li, Joule **2**, 1242 (2018).
[10] J. Jones, et al., Science **353**, 150 (2016).
[11] G. Spezzati, Y. Su, J. P. Hofmann, A. D. Benavidez, A. T. DeLaRiva, J. McCabe, A. K. Datye, and E. J. M. Hensen, ACS Catal. **7**, 6887 (2017).
[12] A. Aitbekova, L. Wu, C. J. Wrasman, A. Boubnov, A. S. Hoffman, E. D. Goodman, S. R. Bare, and M. Cargnello, J. Am. Chem. Soc. **140**, (2018).
[13] H. Jeong, G. Lee, B.-S. Kim, J. Bae, J. W. Han, and H. Lee, J. Am. Chem. Soc. **140**, 9558 (2018).
[14] L. DeRita, et al., Nat. Mater. **18**, 746 (2019).
[15] F. Maurer, J. Jelic, J. Wang, A. Gänzler, P. Dolcet, C. Wöll, Y. Wang, F. Studt, M. Casapu, and J.-D. Grunwaldt, Nat. Catal. **3**, 824 (2020).
[16] W. Yuan, et al., Angew. Chem. Int. Ed. **57**, 16827 (2018).
[17] G. Li, K. Fang, Y. Chen, Y. Ou, S. Mao, W. Yuan, Y. Wang, H. Yang, Z. Zhang, and Y. Wang, J. Catal. **388**, 84 (2020).
[18] R. Li, et al., Nat. Commun. **12**, 1406 (2021).
[19] L. Lin, et al., Nat. Commun. **12**, 6978 (2021).
[20] W.-X. Li, C. Stampfl, and M. Scheffler, Phys. Rev. Lett. **90**, 256102 (2003).
[21] S. Li, et al., ACS Catal. **14**, 1608 (2024).
[22] J. Chen, H. Wang, Z. Wang, S. Mao, J. Yu, Y. Wang, and Y. Wang, ACS Catal. **9**, 5302 (2019).
[23] Y. Fan, F. Wang, R. Li, C. Liu, and Q. Fu, ACS Catal. **13**, 2277 (2023).
[24] M. Moliner, J. E. Gabay, C. E. Kliewer, R. T. Carr, J. Guzman, G. L. Casty, P. Serna, and A. Corma, J. Am. Chem. Soc. **138**, 15743 (2016).
[25] Y. Nishihata, J. Mizuki, T. Akao, H. Tanaka, M. Uenishi, M. Kimura, T. Okamoto, and N. Hamada, Nature **418**, 164 (2002).
[26] M. Cargnello, J. J. D. Jaén, J. C. H. Garrido, K. Bakhmutsky, T. Montini, J. J. C. Gámez, R. J. Gorte, and P. Fornasiero, Science **337**, 713 (2012).
[27] S. Chen, et al., ACS Catal. **11**, 5666 (2021).

[28] V. Muravev, G. Spezzati, Y.-Q. Su, A. Parastaev, F.-K. Chiang, A. Longo, C. Escudero, N. Kosinov, and E. J. M. Hensen, Nat. Catal. **4**, 469 (2021).
[29] B. Hu, et al., Adv. Mater. **34**, 2107721 (2022).
[30] E. M. Slavinskaya, R. V. Gulyaev, A. V. Zadesenets, O. A. Stonkus, V. I. Zaikovskii, Yu. V. Shubin, S. V. Korenev, and A. I. Boronin, Appl. Catal. B: Environ. **166–167**, 91 (2015).
[31] V. Muravev, A. Parastaev, Y. van den Bosch, B. Ligt, N. Claes, S. Bals, N. Kosinov, and E. J. M. Hensen, Science **380**, 1174 (2023).
[32] G. S. Parkinson, Z. Novotny, G. Argentero, M. Schmid, J. Pavelec, R. Kosak, P. Blaha, and U. Diebold, Nat. Mater. **12**, 724 (2013).
[33] D. Jiang, G. Wan, J. Halldin Stenlid, C. E. García-Vargas, J. Zhang, C. Sun, J. Li, F. Abild-Pedersen, C. J. Tassone, and Y. Wang, Nat. Catal. **6**, 618 (2023).
[34] Z. Li, D. Wang, Y. Wu, and Y. Li, Natl. Sci. Rev. **5**, 673 (2018).
[35] S. Wei, et al., Nat. Nanotech. **13**, 856 (2018).
[36] S. Hu and W.-X. Li, Science **374**, 1360 (2021).
[37] J. Yang, et al., Angew. Chem. Int. Ed. **59**, 18522 (2020).
[38] H.-X. Mai, L.-D. Sun, Y.-W. Zhang, R. Si, W. Feng, H.-P. Zhang, H.-C. Liu, and C.-H. Yan, J. Phys. Chem. B **109**, 24380 (2005).
[39] L. Sheng, Z. Ma, S. Chen, J. Lou, C. Li, S. Li, Z. Zhang, Y. Wang, and H. Yang, Chin. J. Catal. **40**, 1070 (2019).

**ACKNOWLEDGEMENTS**

We thank Dr. Ruiyang You in the Center of Electron Microscopy at Zhejiang University for data analysis. We acknowledge the financial support of the National key research and development program (2022YFA1505500), the National Natural Science Foundation of China (51971202, 52025011, 92045301, and 52171019), the Key Research and Development Program of Zhejiang Province (2021C01003), the Zhejiang Provincial Natural Science Foundation of China (LR23B030004, LD19B030001), Shanxi-Zheda Institute of Advanced Materials and Chemical Engineering and the Fundamental Research Funds for the Central Universities.